\documentclass[conference,a4paper]{IEEEtran}
\usepackage{amsmath,epsfig,amssymb,verbatim}
\usepackage{cite,citesort}
\usepackage{graphicx}
\usepackage{bm,bbm}
\usepackage{lipsum}
\usepackage{subfigure}

\def\BibTeX{{\rm B\kern-.05em{\sc i\kern-.025em b}\kern-.08em
   T\kern-.1667em\lower.7ex\hbox{E}\kern-.125emX}}

\newtheorem{Theorem}{Theorem}

\newcommand{\E}{\mathbb{E}}

\begin{document}

\title{The Age of Information of Short-Packet Communications: Joint or Distributed Encoding?}
\author{\IEEEauthorblockN{Zhifeng Tang$^{\ast}$, Nan Yang$^{\ast}$, Parastoo Sadeghi$^{\dagger}$, and Xiangyun Zhou$^{\ast}$}
\IEEEauthorblockA{$^{\ast}$School of Engineering, Australian National University, Canberra, ACT 2600, Australia}
\IEEEauthorblockA{$^{\dagger}$School of Engineering and Information Technology, University of New South Wales, Canberra, ACT 2612, Australia}
\IEEEauthorblockA{Email: zhifeng.tang@anu.edu.au, nan.yang@anu.edu.au, p.sadeghi@unsw.edu.au, xiangyun.zhou@anu.edu.au}

}

\maketitle

\begin{abstract}
In this paper, we analyze the impact of different encoding schemes on the age of information (AoI) performance in a point-to-point system, where a source generates packets based on the status updates collected from multiple sensors and transmits the packets to a destination. In this system, we consider two encoding schemes, namely, the joint encoding scheme and the distributed encoding scheme. In the joint encoding scheme, the status updates from all the sensors are jointly encoded into a packet for transmission. In the distributed encoding scheme, the status update from each sensor is encoded individually and the sensors' packets are transmitted following the round robin policy. To ensure the freshness of packets, the zero-wait policy is adopted in both schemes, where a new packet is immediately generated once the source finishes the transmission of the current packet. We derive closed-form expressions for the average AoI achieved by these two encoding schemes and compare their performances. Simulation results show that the distributed encoding scheme is more appropriate for systems with a relatively large number of sensors, compared with the joint encoding scheme.
\end{abstract}


\IEEEpeerreviewmaketitle

\section{Introduction}
In recent years, real-time applications, such as intelligent transport systems and factory automation, have attracted a wide range of interests. In these applications, timely status updates are critical for accurate monitoring and control \cite{Simsek2016,Li2019}. To reduce transmission latency in real-time applications, short packet communications was considered, due to its unique benefits in delay reduction \cite{Sun2018,Huang2019,yuan2021performance,Sun20181,Li2021}. Moreover, in order to fully characterize the freshness of delivered status information, the concept of age of information (AoI) was introduced as a new and effective performance metric \cite{Kaul2011}. Specifically, AoI is defined as the elapsed time since the last successfully received status was generated by the transmitter, which is a time metric capturing both latency and freshness of transmitted status information.

Since being introduced in \cite{Kaul2011}, the concept of AoI has reaped a wide range of attention and interests. 
The authors of \cite{Kaul2012} studied the average AoI in a first-come-first-served (FCFS) system. Different from \cite{Kaul2012}, \cite{Kaul2012b} proposed a last-come-first-served (LCFS) system, which was shown to achieve a lower average AoI than the FCFS system. Moreover, \cite{Kaul2012b} introduced two different packet management policies for the LCFS system, namely, the preemption policy and the non-preemption policy. In the preemption policy, when a new packet is generated at the source, it is allowed to replace the current packet in service. In contrast, for the non-preemption policy, the newly generated packet has to wait for the current packet in service to finish. In \cite{Yates2012}, the average AoI was analyzed in a multi-user system under the FCFS queuing policy. By considering the impact of unreliable channels on the status transmission loss, \cite{Yates2017} introduced a feedback mechanism to improve the timeliness of the delivered status information. By considering sporadic packet generation rates of users, \cite{Tang2020} proposed a random access based transmission scheme in a multi-user system to improve the average AoI performance. In addition, \cite{sun2019} proposed a Whittle index based scheduling policy to minimize the average AoI, where both a stochastic packet generation model and an unreliable link were considered.

Motivated by the benefits of short packet communications on latency reduction, the AoI performance of short packet communications was analyzed in \cite{devassy2018delay,Devassy2019,Basnay2021,WangGC2019} to evaluate the impact of short packets on the freshness of transmitted information. Specifically, \cite{devassy2018delay} investigated the impact of the packet blocklength on the delay and the peak AoI in a point-to-point communication system. Considering the same system, \cite{Devassy2019} extended \cite{devassy2018delay} to analyze the probability that the peak-age violation exceeds a desired threshold. Focusing on a decode-and-forward relaying system, \cite{Basnay2021} estimated the impact of the packet generation rate, the packet blocklength, and the blocklength allocation factor on the average AoI. Furthermore, \cite{WangGC2019} studied the optimal packet blocklength of non-preemption and preemption policies for minimizing the average AoI. Although the aforementioned studies have investigated the impact of the packet blocklength on the AoI performance of short packet communication systems, the impact of different encoding schemes on the AoI performance has not been touched.

In this paper, we analytically assess the impact of two encoding schemes on the AoI performance of a short packet communication system. In this system, a source transmits status updates received from multiple sensors to a destination, where such status updates can be jointly encoded into a packet for transmission, referred to as the \textit{joint encoding scheme}, or encoded into separate packets individually, referred to as the \textit{distributed encoding scheme}. In these two encoding schemes, the packet generation follows the zero-wait policy, where the source generates a new packet immediately after finishing the transmission of the current packet. We derive new closed-form expressions for the average AoI achieved by both encoding schemes. Moreover, we theoretically determine the condition when one encoding scheme has a better AoI performance than the other. Using simulations, we demonstrate the accuracy of our analytical results and discuss the appropriate choice of 
the encoding scheme to decrease the average AoI based on the information redundancy and the number of sensors.

\section{System Model and Average AoI}\label{Sec:systemmodel}

\begin{figure}[t]
    \centering
   \includegraphics[width=0.65\columnwidth]{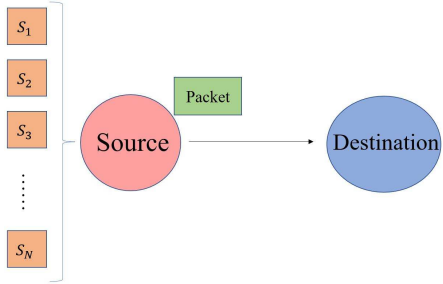}
   \vspace{-1em}
   \centering
    \caption{Illustration of our considered system where a source collects status updates from $N$ sensors and transmits the encoded packets to the destination.}\label{fig:system_model}
    \vspace{-1.5em}
\end{figure}

\subsection{System Description}
We consider a point-to-point wireless communication system as depicted in Fig.~\ref{fig:system_model}, where a source transmits packets to a destination. In this system, the source collects status updates from $N$ sensors and generates packets based on these collected status updates. We denote the $n$th sensor by $S_{n}$, where $n=1,2,\cdots,N$. To ensure the freshness of packets, the zero-wait policy is adopted, where a new packet is immediately generated once the source finishes the transmission of the current packet. In this system, we introduce two encoding schemes, namely, the joint encoding scheme and the distributed encoding scheme.

In the joint encoding scheme, we assume that the joint status update from $N$ sensors at the source contains a fixed number of bits, denoted by $L$. Once the source collects the status update, the source encodes this $L$ bits status update into a packet with the blocklength as $M$ channel use (c.u.) and transmits this packet to the destination. In the distributed encoding scheme, the source collects the status update from sensors and follows a round robin policy. In each time, the source collects the fixed $L_n$ bits status update from the sensor $S_n$ and then encodes the $L_n$ bits status update into a packet with $M_n$ c.u.. Here, we assume that the $N$ sensors are homogeneous sensors, where they have the same length of status updates and blocklength of packets, i.e., $L_n=L_h$ and $M_n = M_h$, $\forall n$. After packet generation, the source transmits the packet to the destination. For the transmission, we define the coding rate, $R$, as the ratio between the number of bits in the status update and the blocklength of the transmitted packet, i.e., $R=\frac{L}{M}$ in the joint encoding scheme and $R=\frac{L_h}{M_h}$ in the distributed encoding scheme.
 
We clarify that the information for the total $N$ sensors are $L$ bits in the joint encoding scheme and $NL_h$ bits in the distributed encoding scheme. According to \cite{Csiszar2011}, the joint information is less than or equal to the sum of the individual information, i.e., $L\leq NL_h$. Thus, we denote the information redundancy over all the sensors by $\alpha$ bits, i.e., $\alpha = NL_h-L$.




As in \cite{WangGC2019}, we assume an additive white Gaussian noise (AWGN) channel between the source and the destination. According to \cite{Polyanskiy2010}, the block error rate for the AWGN channel using finite block length coding can be approximated as


\begin{align}\label{eq:blockerrrate}
    \epsilon(l,m,\gamma) = Q\left(\frac{\frac{1}{2}\log_2(1+\gamma)-\frac{l}{m}}{\log_2(e)\sqrt{\frac{1}{2m}\left(1-\frac{1}{(1+\gamma^2)}\right)}}\right),
\end{align}
where $l$ is the number of bits in the status update, $m$ is the blocklength of the transmitted packet, $\gamma$ is the received signal-to-noise (SNR) ratio at the destination, and $Q(x)=\int_{x}^{\infty}\!\frac{1}{\sqrt{2\pi}}\!\exp\left(\!-\frac{t^2}{2}\!\right)\!\mathrm{d}t$ is the $Q$-function. The expression for the block error rate in \eqref{eq:blockerrrate} is very tight for short packet communications when $m\!\geq\!100$ \cite{Polyanskiy2010} and is close to $0$ for a significantly large $m$. For the packet transmission in our considered system, we denote $T_u$ as the transmission time for each c.u.. Hence, the system can be considered as a time-slotted system where the time duration of each time slot is $T_u$.


\subsection{AoI Formulation}
\begin{figure}[t]
    \centering
    \includegraphics[width=0.8\columnwidth]{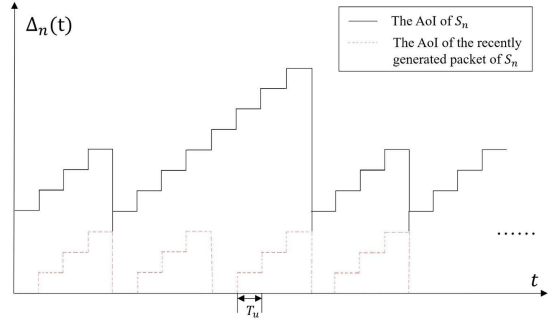}
    \vspace{-1em}
    \centering
    \caption{The AoI variation of $S_n$ in the proposed time-slotted system.}\label{fig:AoIevo}
    \vspace{-1.5em}
\end{figure}
In this subsection, we formulate the expression for the average AoI of the considered time-slotted system where $T_u$ is treated as the unit time.

We denote $\Delta_n(t)$ as the AoI of $S_n$ at time slot $t$. 
Fig.~\ref{fig:AoIevo} plots a sample variation $\Delta_n(t)$ as a function of $t$. We assume that the observation begins at $t=0$ with the AoI of $\Delta_n(0)$. From Fig.~\ref{fig:AoIevo}, we express the AoI at time $t$ as
\begin{align}\label{eq:2}
    \Delta_n(t) = t - u_n(t),
\end{align}
where $u_n(t)$ is the generation time of the most recently received status update at destination from $S_n$ at time $t$. Based on the zero-wait policy, we note that the generation time of the most recently received packet is the time that the source finishes its transmission. We then denote $\Delta(t)$ as the overall AoI of the system at time $t$. Based on \eqref{eq:2}, $\Delta(t)$ is defined as 
\begin{align}\label{eq:AoIDefine}
    \Delta(t) = \max\limits_n\{\Delta_n(t)\},
\end{align}
which is the maximum AoI of all the sensors at time $t$. Then, the average AoI over the observation window of $T$ time slots is calculated as
\begin{align}
    \Delta = \frac{1}{T}\sum\limits_{t=0}^{T-1}\Delta(t). 
\end{align}

We next discuss the AoI expressions for the joint encoding scheme and the distributed encoding scheme separately, due to the fundamental differences in these two schemes.

\begin{enumerate}
    \item In the joint encoding scheme, since the source collects the joint status update from all the sensors, the AoI of different sensors at any time is the same, i.e.,
    \begin{align}
        \Delta(t) = \Delta_{n_1}(t) = \Delta_{n_2}(t),
    \end{align}
    where $n_1$ and $n_2$ $\in\{1,2,\cdots,N\}$. For the convenience of the following analysis, we consider that $T$ time slots are grouped into $K$ frames and each frame has $M$ time slots, i.e., $T=KM$. By averaging $\Delta(t)$ over a sufficiently long observation window with $T\rightarrow\infty$, we obtain the average AoI as
    \begin{align}\label{eq:AoIevoJoint}
        \Delta_J &= \lim_{T\rightarrow\infty}\frac{1}{T}\sum\limits_{t=0}^{T-1}\Delta(t) \notag\\
        &= \lim_{K\rightarrow\infty}\frac{1}{M}\sum\limits_{m=0}^{M-1}\frac{1}{K}\sum\limits_{k=0}^{K-1} \Delta(kM+m).
    \end{align}
   Since $\Delta(kM+m) = \Delta(kM)+m$, we simplify \eqref{eq:AoIevoJoint} as
    \begin{align}\label{eq:jointAoI}
        \Delta_J &= \lim_{K\rightarrow\infty}\frac{1}{M}\sum\limits_{m=0}^{M-1}\frac{1}{K}\sum\limits_{k=0}^{K-1} \left(\Delta(kM)+m\right)\notag\\
        &= \mathrm{AoI}_{0} +\frac{1}{M}\sum\limits_{m=0}^{M-1}m\notag\\
        &= \mathrm{AoI}_{0} + \frac{M-1}{2},
    \end{align}
    where $ \mathrm{AoI}_{0}= \lim_{K\rightarrow \infty}\frac{1}{K}\sum\limits_{k=0}^{K-1} \Delta(kM)$.

\item    In the distributed encoding scheme, we consider $T$ time slots grouped into $K$ frames and each frame has $NM_h$ time slots, i.e., $T=KNM_h$. Then, the average AoI is calculated as
\begin{align}\label{eq:AoIevoDWoutF}
        \Delta_D &= \lim_{T\rightarrow\infty}\frac{1}{T}\sum\limits_{t=0}^{T-1}\Delta(t) \notag\\
        &= \lim_{K\rightarrow\infty}\frac{1}{NM_h}\sum\limits_{m=0}^{NM_h-1}\frac{1}{K}\sum\limits_{k=0}^{K-1} \Delta(kNM_h+m)\notag\\
        &= \lim_{K\rightarrow\infty}\frac{1}{M_h}\sum\limits_{m=0}^{M_h-1}\frac{1}{N}\sum\limits_{n=1}^{N}\frac{1}{K}\sum\limits_{k=0}^{K-1}\Delta(knM_h+m) \notag\\
        &=\frac{1}{M_h}\sum\limits_{m=0}^{M_h-1} \frac{1}{N}\sum\limits_{n=1}^{N}\mathrm{AoI}_{{n,m}},
\end{align}
    where $ \mathrm{AoI}_{{n,m}}=\lim_{K\rightarrow \infty}\frac{1}{K}\sum\limits_{k=0}^{K-1} \Delta(knM_h+m)$. Due to the same length of status updates and blocklength of packets for $N$ sensors, we note that
    \begin{align}\label{eq:AoIn12eq}
        \mathrm{AoI}_{{n_1,m}} = \mathrm{AoI}_{{n_2,m}},
    \end{align}
   for any $n_1$, $n_2\in \{1,2, \cdots, N\}$. Combining \eqref{eq:AoIn12eq} with $\Delta(knM_h+m) = \Delta(knM_h)+m$, we simplify \eqref{eq:AoIevoDWoutF} as
    \begin{align}\label{eq:AoIDis}
       \Delta_D =  \frac{1}{M_h}\sum\limits_{m=0}^{M_h-1} \mathrm{AoI}_{{n,m}} = \mathrm{AoI}_{{N,0}}+\frac{M_h-1}{2}.
    \end{align}
    
 \end{enumerate}   
 

\section{Closed-form Analysis of Average AoI}\label{sec:Derivation}

In this section, we derive the closed-form expressions for the average AoI achieved by the two encoding schemes. We first focus on the joint encoding scheme and present its average AoI in the following theorem.

\begin{Theorem}\label{Theorem:1}
In the joint encoding scheme with the received SNR, $\gamma$, the closed-form expression for the average AoI is derived as
\begin{align}\label{eq:expreAoIpar}
    \Delta_J =\frac{M}{1-\epsilon_J}+\frac{M-1}{2},
\end{align}
where $\epsilon_J=\epsilon(L,M,\gamma)$ and $\epsilon(\cdot,\cdot,\cdot)$ is given in \eqref{eq:blockerrrate}.

\begin{IEEEproof}
In this scheme, based on \eqref{eq:jointAoI}, we need to derive $ \mathrm{AoI}_{0}$ to obtain the average AoI. Note that the AoI expression is periodic with the period $M$, which is given by
\begin{align}\label{eq:aoiende}
{\Delta}\left((k+1)M\right)=
\begin{cases}
M, & \mathcal{C}_k,\\
{\Delta}(kM)+M, & \mbox{otherwise},
\end{cases}
\end{align}
where $\mathcal{C}_k$ is the event that the packet generated at $t=kM$ is successfully received by the destination. By exploiting the probability of the occurrence of $\mathcal{C}_{k}$, denoted by $\textrm{Pr}(\mathcal{C}_{k})$, we obtain
\begin{align}\label{eq:13}
    \mathbb{E} [\Delta((k\!+\!1)kM)]\! =\!\mathrm{Pr}(\mathcal{C}_k)M\!+\! (1\!-\!\mathrm{Pr}(\mathcal{C}_k))\mathbb{E}[\Delta(kM)\!+\!M],
\end{align}
where $\mathrm{Pr}(\mathcal{C}_k) = 1-\epsilon_J$. According to \cite{Yates2019}, we obtain
\begin{align}\label{eq:14}
    \mathrm{AoI}_{0} = \mathbb{E}[{\Delta}(kM)]
=\mathbb{E}[{\Delta}((k+1)M)].
\end{align}
Combining \eqref{eq:14} with \eqref{eq:13}, we derive $ \mathrm{AoI}_{0}$ as
\begin{align}\label{eq:AoI0expression}
    { \mathrm{AoI}_{{0}}}  = \frac{M}{1-\epsilon_J}.
\end{align}
By substituting \eqref{eq:AoI0expression} into \eqref{eq:AoIevoJoint}, we obtain \eqref{eq:expreAoIpar}.
\end{IEEEproof}
\end{Theorem}

We then derive and present the closed-form expression for the average AoI achieved by the distributed encoding scheme in the following theorem.

\begin{Theorem}\label{Theorem:2}
In the distributed transmission scheme with the received SNR, $\gamma$, the closed-form expression for the average AoI is derived as
\begin{align}\label{eq:expreAoIpar2}
    \Delta = \sigma NM_h+\beta M_h+\frac{M_h-1}{2},
\end{align}
where 
\begin{align}
\sigma &= \sum\limits_{n=0}^{N}\binom{N}{n}\frac{(-\epsilon_D)^n}{1-\epsilon_D^n}
\end{align}
and
\begin{align}
    \beta
    =& (1\!-\!\epsilon_D)\sum\limits_{n=1}^{N}(N\!-\!n\!+\!1)\notag\\
    &\times\sum\limits_{n_1=0}^{n-1}\sum\limits_{n_2=0}^{N-n}\binom{n-1}{n_1}\binom{N-n}{n_2}\frac{(-1)^{n_1+n_2}\epsilon_D^{n_2}}{1-\epsilon_D^{n_1+n_2+1}},
\end{align}
with $\epsilon_D = \epsilon(L_h,M_h,\gamma)$ and $\epsilon(\cdot,\cdot,\cdot)$ is given in \eqref{eq:blockerrrate}.

\begin{IEEEproof}
In this scheme, based on \eqref{eq:AoIDis}, we need to derive $ \mathrm{AoI}_{{N,0}}$ to obtain the average AoI. 
Note that the AoI expression is periodic with period $NM_h$ and the AoI of $S_n$ at $t=kNM_h$ is given by
\begin{align}\label{eq:AoIevoDwoutFeachSensor}
    {\Delta}_{n}(kNM_h) = f_nNM_h + (N-n+1)M_h,
\end{align}
where $f_n$ is the number of consecutive transmission failures for packets from $S_n$. We find that $f_n$ follows a geometric distribution, whose probability mass function (PMF) is given by
\begin{align}\label{eq:fneqf}
    \mathrm{Pr}( f_n = f)\! =\! \epsilon_D^{f}(1\!-\!\epsilon_D).
\end{align}
Based on \cite{Yates2019}, we obtain
\begin{align}
    \mathbb{E}[{\Delta}_n(kNM_h)]
=\mathbb{E}[{\Delta}_n((k+1)NM_h)].
\end{align}
According to \eqref{eq:AoIDefine}, \eqref{eq:AoIn12eq}, and \eqref{eq:AoIevoDwoutFeachSensor}, we obtain $ \mathrm{AoI}_{{N,0}}$ in \eqref{eq:AoIDis} as
 \begin{align}\label{eq:AoIcase2d}
     \mathrm{AoI}_{{N,0}} &= \max\limits_{n} \mathbb{E}[{\Delta}_n(kNM_h)]\notag\\
     & = \max\limits_{f_n} \E\left[f_nNM_h + (N-n+1)M_h\right]\notag\\
     &=\E[f_{max}NM_h] + \E\left[(N-n^{\ast}+1)M_h\right],
 \end{align}
 where $f_{max}=\max\{f_n\}$ and $n^{\ast} = \min  \left\{{\arg\max\limits_n f_{max}}\right\}$. It shows that the system AoI is dominated by the largest number of consecutive transmission failures in this scheme. Accordingly, the cumulative distribution function (CDF) of $f_{max}$ is given as
 \begin{align}
     F_{f_{max}}(f) =  \mathrm{Pr}(f_{max}< f) = (1-\epsilon_D^f)^N.
 \end{align}
 Based on the CDF of $f_{max}$, we obtain the PMF of $f_{max}$ as
 \begin{align}\label{eq:PMFfmax}
      \mathrm{Pr}(f_{max}\!=\! f)
      &=\left(1-\epsilon_D^{f+1}\right)^N-\left(1-\epsilon_D^f\right)^N.
 \end{align}
According to the PMF of $f_{max}$ given in \eqref{eq:PMFfmax}, the first item in \eqref{eq:AoIcase2d} is calculated as
\begin{align}\label{eq:AoIcase2part11}
     &\E[f_{max}NM_h] = \sum\limits_{f=0}^{\infty}fNM_h\mathrm{Pr}(f_{max}= f)\notag\\
     &=NM_h\sum\limits_{f=0}^{\infty}f\left(\left(1-\epsilon_D^{f+1}\right)^N-\left(1-\epsilon_D^f\right)^N\right).
\end{align}
We note that
\begin{align}
&\sum\limits_{f=0}^{\infty}f\left(\left(1-\epsilon_D^{f+1}\right)^N-\left(1-\epsilon_D^f\right)^N\right)\notag\\
=&\sum\limits_{f=0}^{\infty}f\sum\limits_{n=0}^{N}\binom{N}{n}\left(\left(-\epsilon_D^{f+1}\right)^n-\left(-\epsilon_D^f\right)^n\right) = \sigma.
\end{align}
Hence, we obtain
\begin{align}\label{eq:AoIcase2part1}
    \E[f_{max}NM_h] = \sigma NM_h.
\end{align}
We then calculate the second item in \eqref{eq:AoIcase2d}. We first note that the number of consecutive transmission failures of the packets from $S_{n^{\ast}}$ is the largest number of consecutive transmission failures of the packets from all the sensors, i.e., $f_{n^{\ast}}=f_{max}$. We also note that the number of consecutive transmission failures of the packets from $S_{n^{\ast}}$ is greater than the number of consecutive transmission failures of the packets from $S_k$, $\forall k<n^{\ast}$, i.e., $f_{n^{\ast}}>f_k$. Thus, we obtain
\begin{align}
    \mathrm{Pr}\left(n^{\ast} =n \right)=& \sum\limits_{f=0}^{\infty}\prod\limits_{k=1}^{n-1}\mathrm{Pr}(f_k<f)\notag\\
    &\times\prod\limits_{k=n+1}^{N}\mathrm{Pr}(f_k\leq f)\mathrm{Pr}(f_n = f).
\end{align}
According to $\mathrm{Pr}(f_k<f) = 1-\epsilon_D^f$, $\mathrm{Pr}(f_k\leq f) = 1-\epsilon_D^{f+1}$, and \eqref{eq:fneqf}, we obtain
\begin{align}
    \mathrm{Pr}\left(n^{\ast}\! =\!n \right)= \sum\limits_{f=0}^{\infty}\left(1\!-\!\epsilon_D^f\right)^{n-1}\left(1\!-\!\epsilon_D^{f\!+\!1}\right)^{N-n}\epsilon_D^f(1\!-\!\epsilon_D).
\end{align}
Thus, the second item in \eqref{eq:AoIcase2d} is calculated as
\begin{align}\label{eq:AoIcase2part2}
    \E\left[(N-n^{\ast}+1)M_h\right]&=\sum\limits_{n=1}^{N}\mathrm{Pr}\left(n^{\ast} =n \right)(N-n+1)M_h\notag\\
    &= \beta M_h.
\end{align}

By substituting \eqref{eq:AoIcase2part1} and \eqref{eq:AoIcase2part2} into \eqref{eq:AoIcase2d}, we obtain the expression for $ \mathrm{AoI}_{{N,0}}$ as given in \eqref{eq:AoIcase2d}. In addition, by substituting the expression for $ \mathrm{AoI}_{{N,0}}$ into \eqref{eq:AoIDis}, we obtain the final result for the distributed encoding scheme.
\end{IEEEproof}
\end{Theorem}

Based on Theorem~\ref{Theorem:1} and Theorem~\ref{Theorem:2}, we obtain the expressions for the average AoI achieved by the joint encoding scheme and the distributed encoding scheme, respectively. Since the block error rate is the function of the blocklength of a packet, there would exist the optimal blocklength of a packet to minimize their average AoI for both encoding schemes. Due to the complexity of the block error rate given in \eqref{eq:blockerrrate}, it is hard to derive closed-form expressions for the optimal blocklength for the proposed encoding schemes. Thus, we resort to numerical methods to find the optimal blocklength, which are the solutions to the equation $\frac{\mathrm{d}\Delta_J}{\mathrm{d}M}=0$ in the joint encoding scheme and the equation $\frac{\mathrm{d}\Delta_D}{\mathrm{d}M_h}=0$ in the distributed encoding scheme.

We then compare the average AoI performance achieved by both encoding schemes by considering the information redundancy $\alpha$. Here, we consider the same coding rate, $R$, in the joint encoding scheme and the distributed encoding scheme, i.e., $R=\frac{L}{M}=\frac{L_h}{M_h}$. We assume that the coding rate leads to a low block error rate in both encoding schemes, which is due to the fact that high reliable communication technologies are typically adopted in practice. Theorem~\ref{Theorem:3} compares the average AoI between two encoding schemes based on the information redundancy.

\begin{Theorem}\label{Theorem:3}
In the system under the assumption of the same coding rate, $R$, and low block error rate for both encoding schemes, the joint encoding scheme has a better AoI performance than the distributed encoding scheme, if $\alpha\geq\alpha_0$, and the distributed encoding scheme has a better AoI performance than the joint encoding scheme, if $\alpha<\alpha_0$. Specifically, $\alpha_0$ is obtained as
\begin{align}
    \alpha_0=\frac{(3-2\sigma)N-2\beta-1}{3}L_h.
\end{align}

\begin{IEEEproof}
 According to \cite{Lopez2011}, the block error rate in the joint encoding scheme can be approximated as $\epsilon_J \approx \epsilon_D^N$. Once $\epsilon_D$ is low, the block error rate $\epsilon_J$ in the joint encoding scheme is negligible. Hence, the average AoI difference between the joint encoding scheme and the distributed encoding scheme is approximated as 
\begin{align}\label{eq:DeltaDf}
    \Delta_{\textrm{Diff}} &= \Delta_{J}-\Delta_D \notag\\
    &\approx\left(\frac{3}{2}\left(N-\frac{\alpha}{L_h}\right)-\left(N\sigma+\frac{1}{2}+\beta\right)\right)M_h.
\end{align}
Based on \eqref{eq:DeltaDf}, we obtain $\Delta_{\textrm{Diff}}\leq 0$ when $\alpha\geq \alpha_0$, and $\Delta_{\textrm{Diff}}> 0$ when $\alpha< \alpha_0$.

\end{IEEEproof}
\end{Theorem}

Theorem~\ref{Theorem:3} reveals that the joint encoding scheme is better when we can compress much information into a joint packet, but the distributed encoding scheme is better when there is low correlation among the information from different sensors.

\section{Numerical Results}\label{sec:Numerical}
In this section, we present numerical results and evaluate the impact of various parameters, including the coding rate, the number of bits in a status update, the number of sensors, and the information redundancy on the average AoI achieved by the two encoding schemes in our considered system.

\begin{figure}[!t]
\centering
\includegraphics[width=0.9\columnwidth]{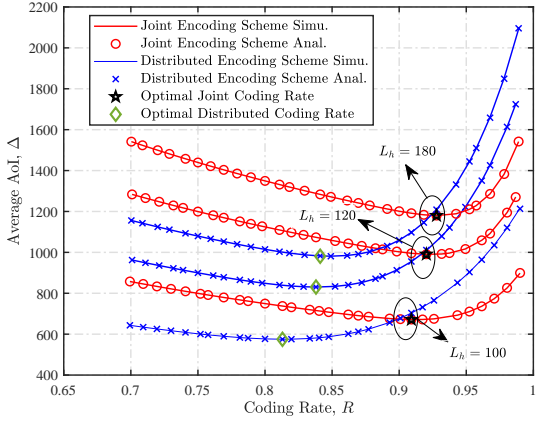}
\vspace{-0.5em}
\caption{The average AoI of the considered system versus the coding rate, $R$, with $N=4$, $\gamma=3$, and $\alpha=0$.}\label{fig:2ps}
\vspace{-1em}
\end{figure}

Fig.~\ref{fig:2ps} plots the average AoI of the considered system versus the coding rate, $R$. We first observe that the analytical average AoI precisely matches the simulation results, which demonstrates the correctness of our analytical results in Theorem~\ref{Theorem:1} and Theorem~\ref{Theorem:2}. We then observe that for both encoding schemes, the average AoI first decreases and then increases when $R$ increases. 
This observation is due to the fact that the increase in $R$ has a two-fold effect on the average AoI via the blocklength of a packet and the block error rate. When $R$ is small, its increase leads to a smaller blocklength of packets, which decreases the average AoI of the considered system. When $R$ exceeds a certain threshold, its increase leads to the significant increase in the block error rate, thereby degrading the AoI performance. We further observe that the optimal coding rate, which minimizes the average AoI, increases when the number of bits in a status update increases in both encoding schemes. This is because that the increase in the number of bits in a status update leads to the increase in the transmission time and the decrease in the block error rate. When the number of bits in a status update is low, the decrease in the block error rate dominantly and positively affects the average AoI. When the number of bits in a status update is high, the increase in the transmission time dominants the average AoI, thereby degrading the AoI performance.   

\begin{figure}[!t]
\centering
\includegraphics[width=0.9\columnwidth]{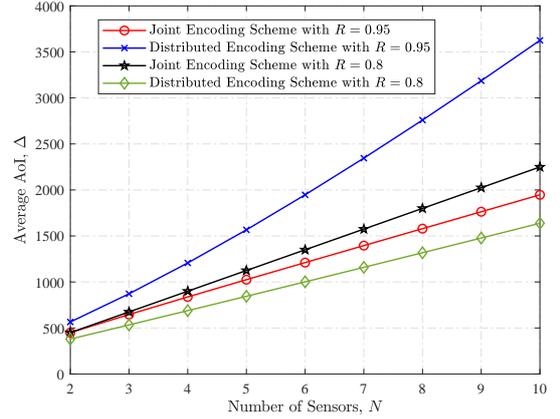}
\vspace{-0.5em}
\caption{The average AoI of the considered system versus the number of sensors, $N$, with $L_h=120$, $\gamma=3$, and $\alpha=0$.}\label{fig:3ps}
\vspace{-1em}
\end{figure}

Fig.~\ref{fig:3ps} plots the average AoI of the considered system versus the number of sensors, $N$. We first observe that when $N$ increases, the average AoI achieved by both encoding schemes increases monotonically. This is because that the increase in $N$ results in the longer transmission time of status updates from all the sensors, which increases the average AoI of the system. We then observe that the joint encoding scheme achieves a lower average AoI than the distributed encoding scheme for a high coding rate, but a larger average AoI for a low coding rate. This is because that the block error rate increases when the coding rate increases. This block error rate has a more pronounced impact on the average AoI for the distributed encoding scheme than the joint encoding scheme. 

\begin{figure}[!t]
\centering
\includegraphics[width=0.9\columnwidth]{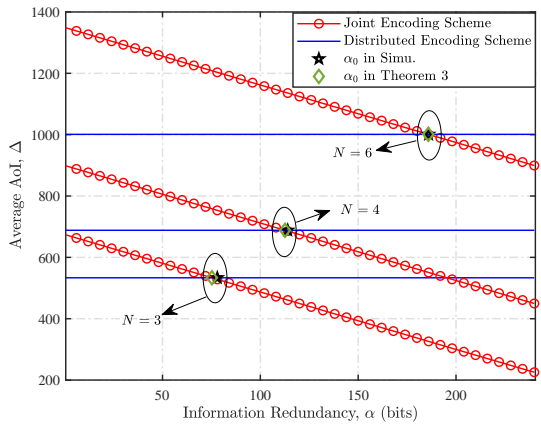}
\vspace{-0.5em}
\caption{The average AoI of the considered system versus the information redundancy, $\alpha$, with $L_h=120$, $\gamma=3$, and $R=0.8$.}\label{fig:4ps}
\vspace{-1em}
\end{figure}

Fig.~\ref{fig:4ps} plots the average AoI of the considered system versus the information redundancy, $\alpha$. We first observe that the analytical $\alpha_0$ tightly matches the simulation result, which demonstrates the correctness of our analytical result in Theorem~\ref{Theorem:3}. It indicates that we can adopt Theorem \ref{Theorem:3} to find the optimal encoding scheme to minimize the average AoI based on the information redundancy. We then observe that the average AoI decreases when $\alpha$ increases for the joint encoding scheme for different $N$. This is because that the increase in $\alpha$ leads to the small number of bits in status updates for the transmission in the joint encoding scheme. Hence, the blocklength of a packet for the transmission decreases, which decreases the average AoI of the system. We further observe that $\alpha_0$ increases as the number of sensors increases. This observation is due to the fact that when $N$ is large, the distributed encoding scheme significantly increases the update frequency of the states received at the destination, compared with the joint encoding scheme, which decreases the average AoI. This observation also implies that it is better to select the joint encoding scheme when the number of sensors is small, but select the distributed encoding scheme when the number of sensors is large.

\section{Conclusion}\label{sec:Conclusion}
This paper considered a point-to-point system where a source generates packets and transmits them to a destination under an unreliable channel. We analyzed the average AoI achieved by two encoding schemes, i.e., the joint encoding scheme and the distributed encoding scheme, under a finite packet block length model. By using simulation results, we demonstrated the accuracy of our analysis 
and showed that the block error rate has a more significant impact on the average AoI for the distributed encoding scheme than the joint encoding scheme. In addition, we found that the distributed encoding scheme has a better AoI performance than the joint encoding scheme in a highly reliable communication system with a large number of sensors.

\section*{Acknowledgment}

This work was funded by the Australian Research Council Discovery Project DP180104062.

\bibliographystyle{IEEEtran}
\bibliography{bibli.bib}

\begin{thebibliography}{10}
\providecommand{\url}[1]{#1}
\csname url@samestyle\endcsname
\providecommand{\newblock}{\relax}
\providecommand{\bibinfo}[2]{#2}
\providecommand{\BIBentrySTDinterwordspacing}{\spaceskip=0pt\relax}
\providecommand{\BIBentryALTinterwordstretchfactor}{4}
\providecommand{\BIBentryALTinterwordspacing}{\spaceskip=\fontdimen2\font plus
\BIBentryALTinterwordstretchfactor\fontdimen3\font minus
  \fontdimen4\font\relax}
\providecommand{\BIBforeignlanguage}[2]{{%
\expandafter\ifx\csname l@#1\endcsname\relax
\typeout{** WARNING: IEEEtran.bst: No hyphenation pattern has been}%
\typeout{** loaded for the language `#1'. Using the pattern for}%
\typeout{** the default language instead.}%
\else
\language=\csname l@#1\endcsname
\fi
#2}}
\providecommand{\BIBdecl}{\relax}
\BIBdecl

\bibitem{Simsek2016}
M.~{Simsek}, A.~{Aijaz}, M.~{Dohler}, J.~{Sachs}, and G.~{Fettweis},
  ``{5G}-enabled tactile {I}nternet,'' \emph{IEEE J. Select. Areas Commun.},
  vol.~34, no.~3, pp. 460--473, Mar. 2016.

\bibitem{Li2019}
C.~{Li}, N.~{Yang}, and S.~{Yan}, ``Optimal transmission of short-packet
  communications in multiple-input single-output systems,'' \emph{IEEE Trans.
  Veh. Technol.}, vol.~68, no.~7, pp. 7199--7203, Jul. 2019.

\bibitem{Sun2018}
X.~{Sun}, S.~{Yan}, N.~{Yang}, Z.~{Ding}, C.~{Shen}, and Z.~{Zhong}, ``Downlink
  {NOMA} transmission for low-latency short-packet communications,'' in
  \emph{Proc. IEEE Int. Commun. Conf.}, Kansas City, MO, May 2018, pp. 1--6.

\bibitem{Huang2019}
X.~{Huang} and N.~{Yang}, ``On the block error performance of short-packet
  non-orthogonal multiple access systems,'' in \emph{Proc. IEEE Int. Commun.
  Conf.}, Shanghai, China, May 2019, pp. 1--7.

\bibitem{yuan2021performance}
L.~Yuan, Z.~Zheng, N.~Yang, and J.~Zhang, ``Performance analysis of
  short-packet non-orthogonal multiple access with {Alamouti} space-time block
  coding,'' \emph{IEEE Trans. Veh. Technol.}, vol.~70, no.~3, pp. 2900--2905,
  Mar. 2021.

\bibitem{Sun20181}
X.~{Sun}, S.~{Yan}, N.~{Yang}, Z.~{Ding}, C.~{Shen}, and Z.~{Zhong},
  ``Short-packet downlink transmission with non-orthogonal multiple access,''
  \emph{IEEE Trans. Wireless Commun.}, vol.~17, no.~7, pp. 4550--4564, July
  2018.

\bibitem{Li2021}
C.~Li, C.~She, N.~Yang, and T.~Q.~S. Quek, ``Secure transmission rate of short
  packets with queueing delay requirement,'' \emph{IEEE Trans. Wireless
  Commun.}, pp. 1--1, Jul. 2021.

\bibitem{Kaul2011}
S.~{Kaul}, M.~{Gruteser}, V.~{Rai}, and J.~{Kenney}, ``Minimizing age of
  information in vehicular networks,'' in \emph{Proc. IEEE Conf. Sensor Ad Hoc
  Commun. Netw.}, Salt Lake City, UT, Jun. 2011, pp. 350--358.

\bibitem{Kaul2012}
S.~{Kaul}, R.~{Yates}, and M.~{Gruteser}, ``Real-time status: How often should
  one update?'' in \emph{Proc. IEEE Int. Conf. Comput. Commun.}, Orlando, FL,
  Mar. 2012, pp. 2731--2735.

\bibitem{Kaul2012b}
S.~K. {Kaul}, R.~D. {Yates}, and M.~{Gruteser}, ``Status updates through
  queues,'' in \emph{Proc. Conf. on Inf. Sciences and Systems}, Baltimore, MD,
  Mar. 2012, pp. 1--6.

\bibitem{Yates2012}
R.~D. {Yates} and S.~{Kaul}, ``Real-time status updating: Multiple sources,''
  in \emph{Proc. IEEE Int. Sympos. Inf. Theory}, Cambridge, MA, Jul. 2012, pp.
  2666--2670.

\bibitem{Yates2017}
R.~D. {Yates} and S.~K. {Kaul}, ``Status updates over unreliable multiaccess
  channels,'' in \emph{Proc. IEEE Int. Sympos. Inf. Theory}, Aachen, Germany,
  Jun. 2017, pp. 331--335.

\bibitem{Tang2020}
Z.~{Tang}, Z.~{Sun}, N.~{Yang}, and X.~{Zhou}, ``Age of information of
  multi-source systems with packet management,'' in \emph{Proc. IEEE Int.
  Commun. Conf.}, Dublin, Ireland, Jun. 2020, pp. 1--6.

\bibitem{sun2019}
J.~Sun, Z.~Jiang, B.~Krishnamachari, S.~Zhou, and Z.~Niu, ``Closed-form
  {Whittle’s} index-enabled random access for timely status update,''
  \emph{IEEE Trans. Commun.}, vol.~68, no.~3, pp. 1538--1551, Mar. 2020.

\bibitem{devassy2018delay}
R.~Devassy, G.~Durisi, G.~C. Ferrante, O.~Simeone, and E.~Uysal-Biyikoglu,
  ``Delay and peak-age violation probability in short-packet transmissions,''
  in \emph{Proc. IEEE Int. Sympos. Inf. Theory}, Vail, CO, Jun. 2018, pp.
  2471--2475.

\bibitem{Devassy2019}
R.~{Devassy}, G.~{Durisi}, G.~C. {Ferrante}, O.~{Simeone}, and E.~{Uysal},
  ``Reliable transmission of short packets through queues and noisy channels
  under latency and peak-age violation guarantees,'' \emph{IEEE J. Select.
  Areas Commun.}, vol.~37, no.~4, pp. 721--734, Apr. 2019.

\bibitem{Basnay2021}
C.~M.~W. Basnayaka, D.~N.~K. Jayakody, T.~D.~P. Perera, and M.~V. Ribeiro,
  ``Age of information in an {URLLC}-enabled decode-and-forward wireless
  communication system,'' in \emph{Proc. IEEE Veh. Techn. Conf.}, Helsinki,
  Finland, Apr. 2021, pp. 1--6.

\bibitem{WangGC2019}
R.~Wang, Y.~Gu, H.~Chen, Y.~Li, and B.~Vucetic, ``On the age of information of
  short-packet communications with packet management,'' in \emph{Proc. IEEE
  Global Commun. Conf.}, Waikoloa, HI, Dec. 2019, pp. 1--6.

\bibitem{Csiszar2011}
I.~Csiszar and J.~Korner, \emph{Information Theory: Coding Theorems for
  Discrete Memoryless Systems}.\hskip 1em plus 0.5em minus 0.4em\relax
  Cambridge: Cambridge University Press, 2011.

\bibitem{Polyanskiy2010}
Y.~Polyanskiy, H.~V. Poor, and s.~Verdu, ``Channel coding rate in the finite
  blocklength regime,'' \emph{IEEE Trans. Inf. Theory}, vol.~56, no.~5, pp.
  2307--2359, Apr. 2010.

\bibitem{Yates2019}
R.~D. {Yates} and S.~K. {Kaul}, ``The age of information: Real-time status
  updating by multiple sources,'' \emph{IEEE Trans. Inf. Theory}, vol.~65,
  no.~3, pp. 1807--1827, Mar. 2019.

\bibitem{Lopez2011}
M.~Lopez-Benitez and F.~Casadevall, ``Versatile, accurate, and analytically
  tractable approximation for the {G}aussian {Q}-function,'' \emph{IEEE Trans.
  Commun.}, vol.~59, no.~4, pp. 917--922, Feb. 2011.

\end{thebibliography}

\end{document}